\documentclass[sigconf,9pt]{acmart}
\copyrightyear{2024} 
\acmYear{2024} 
\setcopyright{rightsretained} 
\acmConference[NAIC '24]{SIGCOMM Workshop on Networks for AI Computing}{August 4--8, 2024}{Sydney, NSW, Australia}
\acmBooktitle{SIGCOMM Workshop on Networks for AI Computing (NAIC '24), August 4--8, 2024, Sydney, NSW, Australia}
\acmDOI{10.1145/3672198.3673797}
\acmISBN{979-8-4007-0713-1/24/08}



\usepackage{tikz}
\usepackage{amsmath}
\usepackage{algorithmic}
\usepackage{filecontents}

\usepackage{amsmath}

\usepackage{tikz}
\usepackage{xcolor}
\usepackage{xspace}
\usepackage{tikz}
\usepackage{pifont}
\usepackage[english]{babel}
\usepackage{blindtext}
\usepackage{lipsum}
\usepackage[ruled,vlined,linesnumbered]{algorithm2e}
\usepackage{booktabs}

\usepackage{filecontents}
\usepackage{xspace}
\usepackage{subfig}
\usepackage{xcolor}
\usepackage{hhline}
\usepackage{multirow}
\usepackage{soul}

\usepackage{array}
\usepackage{comment}
\usepackage{listings}
\usepackage{dsfont}
\usepackage{xurl}
\usepackage{caption}
\usepackage{ltablex}

\usepackage{ifluatex}

\usepackage{outlines}
\newcommand{\name}{Eloquent\xspace}

\newcounter{packednmbr}
\newenvironment{packedenumerate}{\begin{list}{\thepackednmbr.}{\usecounter{packednmbr}\setlength{\itemsep}{0.5pt}\addtolength{\labelwidth}{-4pt}\setlength{\leftmargin}{2ex}\setlength{\listparindent}{\parindent}\setlength{\parsep}{1pt}\setlength{\topsep}{0pt}}}{\end{list}}
\newenvironment{packeditemize}{\begin{list}{$\bullet$}{\setlength{\itemsep}{0.5pt}\addtolength{\labelwidth}{-4pt}\setlength{\leftmargin}{2ex}\setlength{\listparindent}{\parindent}\setlength{\parsep}{1pt}\setlength{\topsep}{2pt}}}{\end{list}}

\newcommand{\tightcaption}[1]{\vspace{-0.15cm}\caption{{\normalfont{\textit{{#1}}}}}\vspace{-0.3cm}}

\newcommand{\mypara}[1]{\vspace{0.05cm}\noindent{\bf {#1}:}~}

\definecolor{backcolour}{rgb}{0.96,0.96,0.96}
\definecolor{codegray}{rgb}{0.5,0.5,0.5}
\definecolor{deepblue}{rgb}{0,0,0.6}
\definecolor{deepred}{rgb}{0.6,0,0}
\definecolor{deepgreen}{rgb}{0,0.5,0}
\lstdefinestyle{mystyle}{
    backgroundcolor=\color{backcolour},   
    commentstyle=\color{codegreen},
    morekeywords={self, True},
    keywordstyle=\color{deepblue},
    numberstyle=\tiny\color{codegray},
    emph={MyClass,__init__,EncodingType,Image},
    emphstyle=\color{deepred},
    stringstyle=\color{deepgreen},
    basicstyle=\ttfamily\footnotesize,
    breakatwhitespace=false,         
    breaklines=true,                 
    captionpos=b,                    
    keepspaces=true,                 
    numbers=left,                    
    numbersep=5pt,                  
    showspaces=false,                
    showstringspaces=false,
    showtabs=false,                  
    tabsize=1
}

\usepackage{subcaption}

\title{\name: A More Robust Transmission Scheme for LLM Token Streaming}

\begin{document}
% \pagenumbering{gobble}
% \author{Paper \#166}

\author{Hanchen Li, Yuhan Liu, Yihua Cheng, Siddhant Ray,  Kuntai Du, Junchen Jiang \\
\textit{University of Chicago}}
\pagenumbering{gobble}
\begin{abstract}
To render each generated token in real-time for users, the Large Language Model (LLM) server generates tokens one by one and streams each token (or group of a few tokens) through the network to the user right after generation, which we refer to as {\em LLM token streaming}. 
However, under unstable network conditions, the LLM token streaming experience could suffer greatly from stalls since one packet loss could block the rendering of later tokens even if the packets containing them arrive on time.
With a measurement study, we show that current applications suffer from increased stalls under unstable networks. 

For this emerging token streaming problem in LLM Chatbots that differs from previous multimedia and text applications, we propose a novel transmission scheme, called {\em Eloquent}, which puts newly generated tokens as well as currently unacknowledged tokens in the next outgoing packet.
This ensures that each packet contains some new tokens and, in the meantime, is independently rendered when received, avoiding the aforementioned stalls caused by missing packets.
Through simulation under various networks, we show \name reduces stall ratio (proportion of token rendering wait time) by 71.0\% compared to the retransmission method commonly used by real chatbot applications and by 31.6\% compared to the baseline packet duplication scheme.
By tailoring \name to fit the token-by-token generation of LLM, we enable the Chatbots to respond like an eloquent speaker for users to better enjoy pervasive AI.

\end{abstract}
\maketitle
% {\par\smallskip\small\noindent{\bfseries ACM Reference Format:}\par\nobreak
%   \noindent Hanchen Li, Yuhan Liu, Yihua Cheng, Siddhant Ray,  Kuntai Du, Junchen Jiang. 2024. Eloquent: A More Robust Transmission Scheme for
% LLM Token Streaming
%   In \textit{NAIC '24, August 4, 2024, Sydney, Australia. ACM, New York, NY, USA, 7 pages}
%   }

\pagestyle{plain}

\section{Introduction}
Large Language Models (LLMs) are growing more and more popular~\cite{llm1, llm2, llm3}. One of the most important use cases is the Chatbots that utilize LLM models to take in user prompts and send back real-time responses to answer user queries.~\cite{chatgpt, claude, bard}. Significant effort has been made in the machine learning and system community to generate better responses more efficiently~\cite{sheng2023flexgen, kwon2023efficient, dao2022flashattention, 280922, wu2023fast, liu2024andes}. However, the {\em streaming} of these generated tokens to users is also critical. Having intermittent stalls during transmission could break the smoothness of interaction and damage user experience.
% Quality of Experience (QoE) of users.

In a token streaming pipeline of LLM Chatbot, tokens are generated in the cloud data center after the data center receives the prompt. After each token is generated, they are sent over the wide-area network to reach the client device. If some packets are dropped or significantly delayed in the network, the client will need to wait for packets to be retransmitted from the sender. 
This could cause an undesired stall for token streaming since all later tokens need to wait for previous ones to arrive at the receiver before being rendered.

\begin{figure}
\centering
     \includegraphics[width=.99\linewidth]{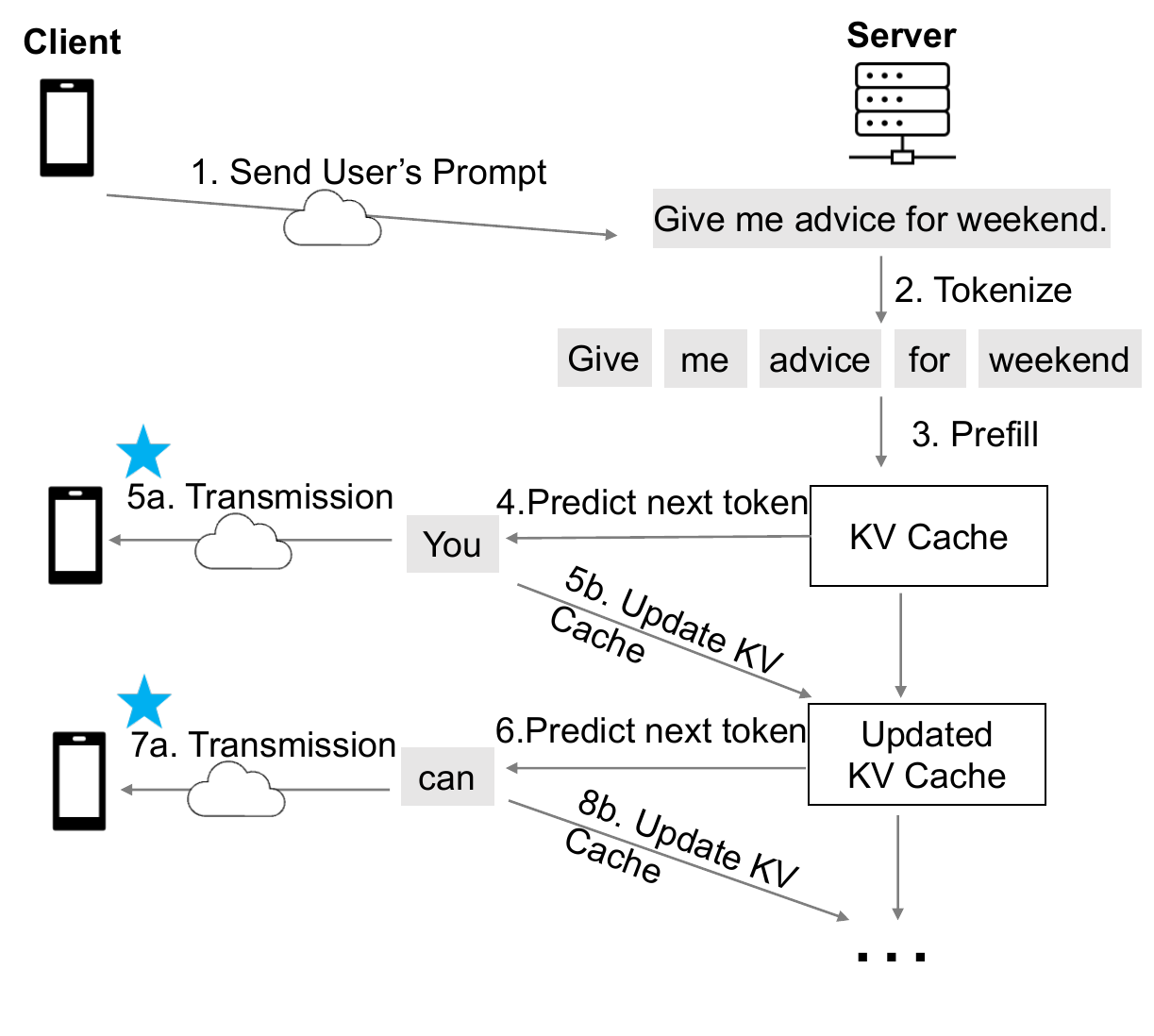}
     \vspace{-12pt}
    \tightcaption{LLM Chatbot Token Streaming Pipeline. We aim to improve the transmission part (Blue Starred 5a, 7a)} 
    \vspace{-2pt}
    \label{fig:pipeline}
\end{figure}

\begin{figure*}[t!]
\centering
     \includegraphics[width=\linewidth]{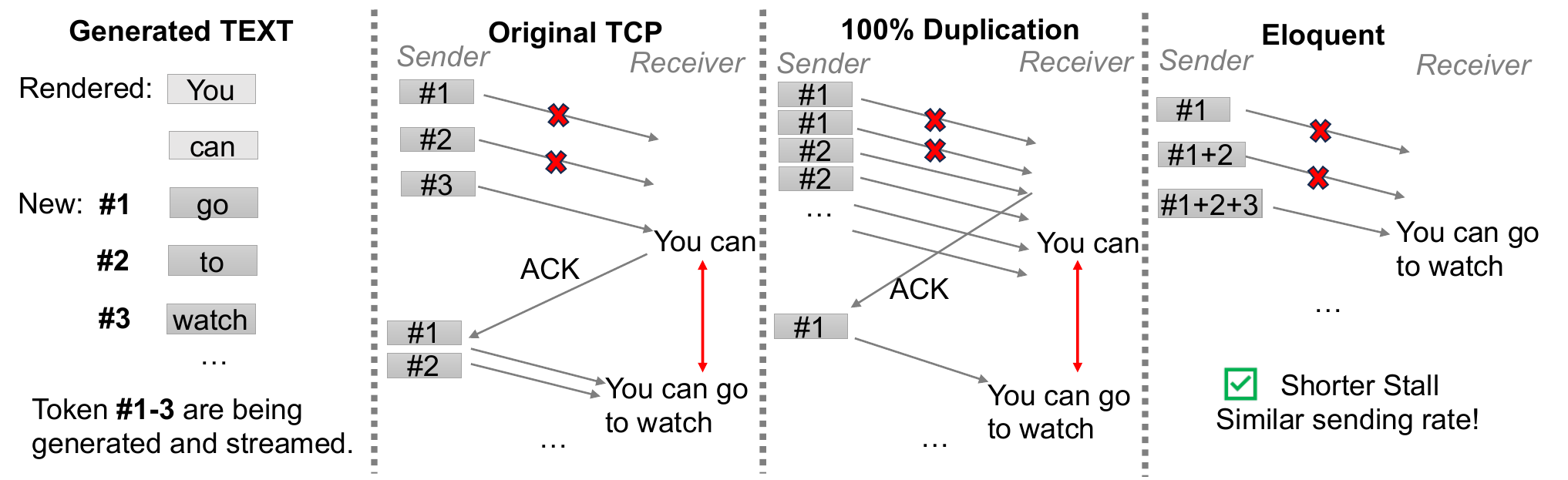}
     \vspace{-10pt}
    \tightcaption{An illustrative example. When the packets containing the first two tokens are lost, \name significantly reduces stall with the similar overall sending rate, compared to TCP or duplicating each packet twice.
    We use \#1, \#2, \#3 to represent the newly generated tokens in later columns and only show  ACKs that could signal retransmission.} 
    \label{fig:comparison}
\end{figure*}

As AI services become ubiquitous, people could use these LLM Chatbot services anywhere, including places where intermittent connections happen, such as on railway~\cite{9131751, 285058} or using a lossy wifi link~\cite{295691}. To observe how current applications perform under such scenarios, we conducted a measurement study to understand how an unstable network affects ChatGPT Streaming API. 
By reading the packet traces, we see that these applications rely on retransmission to handle packet loss, which aggravates the long stall. Moreover, we see an interesting pattern---after some packets are lost, the subsequent packets containing newer tokens may arrive earlier than the retransmission packet. Even so, no more new tokens could be rendered as they are blocked by the lost tokens. Every token must still wait for retransmission before rendering. However, under an unstable network, the Round Trip Time (RTT) could be long~\cite{rtt_long}. This means that all retransmission mechanisms in TCP will suffer from a long response delay, and this causes stall~\cite{rfc793, webrtc}. A natural solution is to add redundancy. 
Unfortunately, sending more duplicated packets together with the original token could fail if the connection was lossy during that time. Sending Forward Error Correction (FEC) packets at the end of a token group will only help after FEC packets arrive at the end, which could already be too late.

Although text streaming is not entirely new, but token streaming {\em differs} in a few key aspects. 
For instance, collaborative editors (like Overleaf or Google Docs) also stream keystrokes from users.
However, compared to LLM’s token generation speed, the typical rate at which users manually type words is arguably slower~\cite{typing_speed}, making TCP retransmission a sensible solution to handle packet loss without backlogging words on the sender side. 
On the other hand, virtual assistants do automatically generate responses. Still, until recently, their text generation models were fast enough to produce the entire response before sending the response to the user as a whole~\cite{siri}.
In contrast, LLMs generate tokens {\em autoregressively}, so every token has to be sent out to users as soon as possible in order to ensure a smooth user experience.

% Moreover, although there have also been similar applications, {\em token streaming} brings a different problem since the tokens are iteratively generated with speed similar to human reading. The outputs from keystrokes for remote shell could be easily predicted~\cite{180894}, while predicting the next token is just the task of LLM. Cooperative document editing has much lower word rate due to human typing speed limit~\cite{typing_speed}. Virtual assistants like Siri only reveal content after the whole response has been received since generation is faster~\cite{siri}. 

To tackle this new {\em token streaming} problem under unstable networks, we proposed a novel transmission scheme, \name, that provides a smooth token streaming experience and transforms Chatbot into a more eloquent speaker under various network conditions by adding smart redundancy.
Demonstrated in Fig \ref{fig:comparison}, \name adds {\em unacked tokens}, defined as tokens whose packets have been sent but have not been ACK'ed by the receiver, into the packet of the newly generated token. It prepares each packet in such a way that each received packet contains sufficient information for the rendering of the new token contained inside, preventing the retransmission blocking issue discussed above.

Through our simulation under various network conditions, \name can reduce stall ratio\footnote{Proportion of token wait time when inter-token gap exceeds 200ms} by 71.0\% compared to the retransmission method used by TCP, and by 31.6\% with less total data sent than baseline packet duplication method. 

% To summarize, our main contributions are to:
% \begin{itemize}
%     \item Identify the transmission problems in  LLM {\em token streaming} pipeline and call for improvement.

%     \item Conduct real-world measurements to show stall increase of current applications under unstable network.

%     \item Propose the design of \name, a novel transport layer scheme
%     tailored for token streaming to reduce stall rates on the client side.

%     % \item Implemented a new simulation platform that decouples token generation in the cloud with transmission for future research in token streaming.   
% \end{itemize}

Admittedly, there remain challenges in deploying \name inside real-world protocols like RTP or QUIC. We hope this work sparks more discussion around taking system components outside datacenters into consideration in the LLM serving pipeline. For this end, we list several future research directions in \S\ref{sec:discussion} along this path.
\section{Background and Motivation}
\subsection{Token Streaming}

We provide a brief end-to-end overview of the current LLM Chatbot streaming pipeline in Fig. \ref{fig:pipeline}. First, the user enters a prompt, which is then sent to the cloud data center. In the data center, the prompt is split into tokens (partial or full words). These tokens then go through a prefill phase to process the prompt in a parallel fashion. 
After the prefill phase, the model iteratively generates the next token with the current sentence, appends the new token to the sentence, and repeats the process until an end-of-sentence (EOS) token is generated. In this output phase, the latest generated tokens are sent to the user through the network right after the generation, rather than waiting for all tokens to be generated, in order to provide a real-time streaming experience. The tokens are rendered on the screen in the order that they are generated. In this paper, we focus on the transmission step of {\em token streaming} under unstable network conditions. 

\subsection{Application Measurement}

As LLM Chatbots are becoming more pervasive in our lives, it is important that the services deliver smooth experiences under different network conditions. However, as previous research and reports~\cite{wireless_1, wireless_2,9131751, 285058, 295691} have pointed out, intermittent connection could happen in many real-world scenarios such as on railway or using a lossy wi-fi link. In these cases, groups of packet might get lost during a short period of time due to a physical surface block or signal interference from other devices. 

To learn about how current LLM Chatbot applications perform under these unstable network conditions, we conducted a measurement study on one of the most popular tools to build LLM Chatbot, ChatGPT API.

\begin{figure}
\centering
     \includegraphics[width=.99\linewidth]{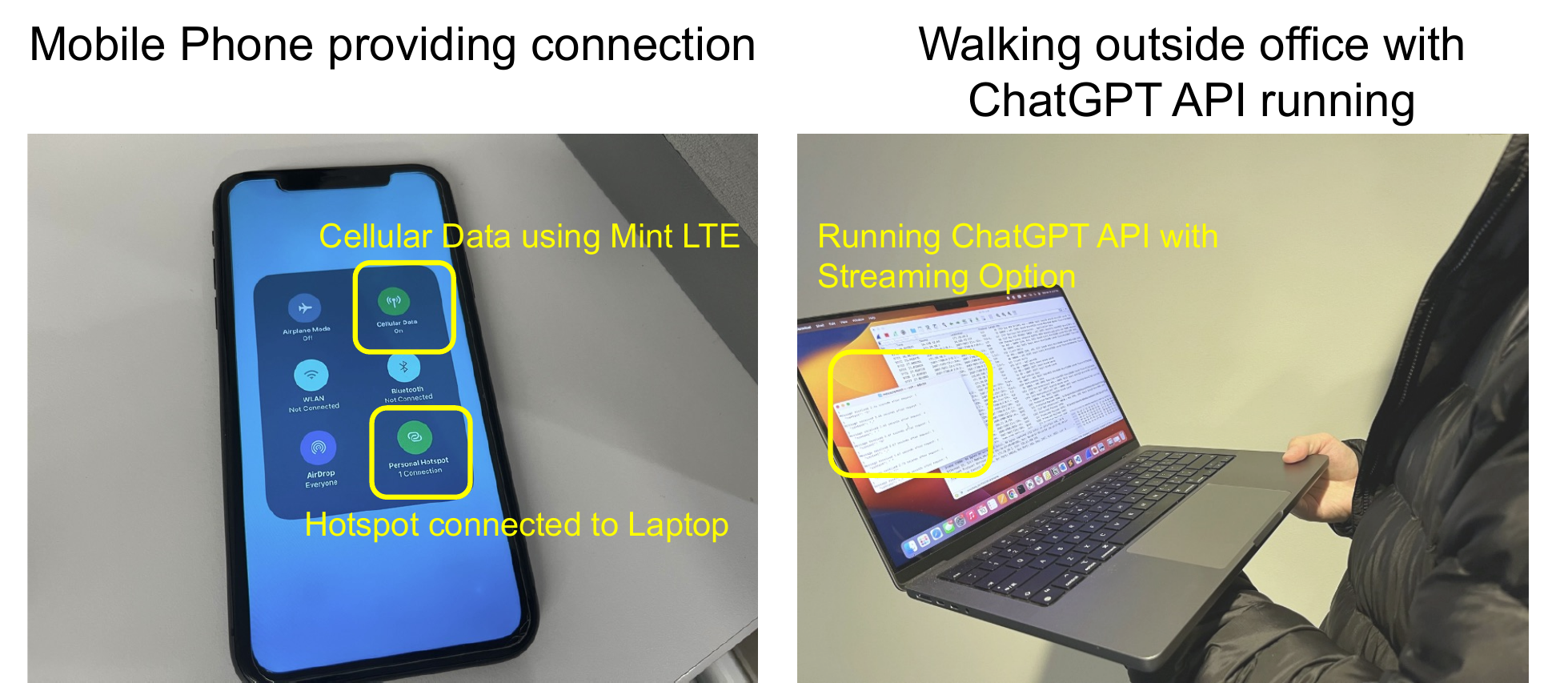}
    \tightcaption{Our Network Measurement Testbed} 
    \label{fig:emulation}
    % \vspace{-8pt}s
\end{figure}

\begin{figure}
\centering
     \includegraphics[width=.99\linewidth]{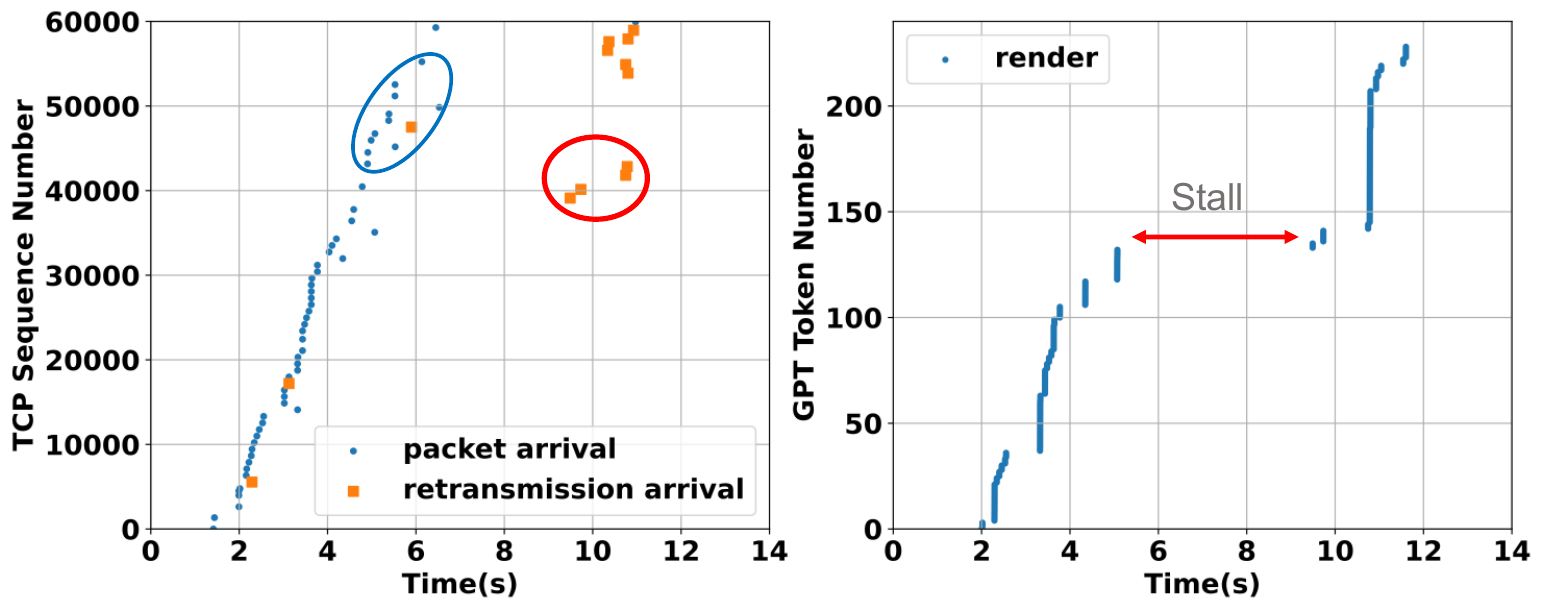}
    \tightcaption{Packet arrival and token rendering of one measured session. Retransmission packets blocking rendering are in circled in red. Blocked packets circled in blue.} 
    \label{fig:api}
    % \vspace{-8pt}
\end{figure}

\mypara{Measurement Testbed}
\label{emulated_measurement}
We emulate an unstable network by connecting a Macbook 2023 M2 laptop to the hotspot of iPhone 11 using Mint LTE service in North America as shown in Fig. ~\ref{fig:emulation}. We put the phone inside an office and walked around outside the office (within 30m). Meanwhile, on the laptop, we run an automated bash script that repetitively starts 30 GPT API sessions using the same prompt with the ``streaming'' option enabled. The prompt was set to ``Count to 80, with a comma between each number and no newlines. E.g., 1, 2, 3, $\cdots$'' for controlled outputs. 
We record packet arrivals with tcpdump~\cite{tcpdump} with the interface flag set to ``any''. To record token arrival times, we slightly modified the sample Python script provided in OpenAI documentation~\cite{openai_example}.

 \mypara{Findings} 
 The packet trace shows that the data is transmitted through encrypted TLS protocol (over TCP).
 Performance-wise, under this unstable network condition, compared with the reference of connecting laptop directly to high-speed wifi in the office, the 95\% token rendering stall\footnote{We defined stall as inter-token gaps larger than 200ms. This follows from works in video streaming~\cite{stall, cheng2023grace}.} increased from 377ms to 3483ms and from 1501.4ms to 2924.4ms, respectively for model "GPT-3.5-turbo" and "GPT-4", under an average network loss rate of 10.5\% and 15.8\%.  

We provide a detailed analysis of a single trace to show why there is potential for 
improvement in Fig. \ref{fig:api}. As the packet arrival graphs show on the left, some packets (red-circled) were lost and retransmitted. However, many packets containing later tokens arrived before the retransmission packets, which arrived much later at around the 9th second due to the unstable connection. 
The impact of this ``retransmission blocking future token rendering'' is severe, as shown on the right-hand side. No tokens were rendered from the moment these packets were lost until the retransmission came through in the 9th second. While many packets arrived after the loss, all of the tokens contained inside were blocked by the lost packets. Thus, a stall lasted almost 5 seconds purely due to the transmission glitch, even though new tokens kept arriving during that period.

% \mypara{Other Applications}
% We extended our measurement to other popular applications and reported the qualitative results. Bard~\cite{bard} maintains a buffer to store arrived tokens and renders them after the buffer contains enough tokens to stream smoothly. This reduces intermittent stall in the middle but sacrifices real-timeness and increases Time to First Token, which damages user experience as well~\cite{ttft1, ttft2}. Claude~\cite{claude} and ChatGPT web version~\cite{chatgpt} both suffer from stall increase in both frequency and length. Although ChatGPT website version and Bard uses QUIC protocol (on UDP), the loss of token is still handled by retransmission.

\subsection{Previous Work}
\label{subsec:previous}
Although the problem of token streaming is new, there have been prior works that aim to protect transmission against packet loss for other media forms.

The most popular method for loss-resilience is Forward Error Correction (FEC)~\cite{tambur, wicker1999reed, mackay2005fountain}. There are different types of coding methods~\cite{wicker1999reed, mackay2005fountain,mackay1997near}, and the fundamental idea is to encode an a-byte segment to (a+b) bytes such that when any a of the (a+b) bytes arrive, the receiver can reconstruct the original data. 
% More recently, people have proposed streaming code that tries to handle burst losses and satisfy latency constraints for streaming applications~\cite{streaming_code, streaming_code2}. However, it is worth noting that the latency requirement for streaming code requires a ``guard space'' of guaranteed arrival after a packet loss period.
Moreover, FEC protects a file (video/audio frame) that spans multiple packets to recover partially delivered data. In token streaming, the minimal transmission unit is one single token, which is far less than the maximal size of one packet.\footnote{Single token packet for Bard is 50-80 bytes; ChatGPT packet is 842 bytes for three tokens.} There have been works that target intermittent connectivity for text-based applications including Mosh for interactive remote shell~\cite{180894}. Mosh predicts output of keystrokes at client-side before the server sends responses back. However, these user keystrokes are sent to the server by clients in the first place, and their output is easy to predict (the output of typing h,e,l,l,o is simply showing “hello” on the terminal). In token streaming, the texts are generated by the server-side LLM and sent back to the client and could hardly be predicted by the client.Cooperative document editor (ex. Google doc) is a similar application. But while LLM generation speed is ~10 tokens/s~\cite{measurement_gpt}, human typing speed is 30-45 words/min~\cite{typing_speed}. This makes online editing already have long gaps (
\char`\~1s) between tokens. Virtual Assistants like Siri utilizes multi-path TCP~\cite{siri} and only reveal content after the whole response has been received, which poses stronger assumption on client device and potentially leads to high time to first token (TTFT) if adapted to LLM Chatbots.

From the LLM systems , although there have been extensive work on scheduling~\cite{wu2023fast, 280922}, memory management~\cite{sheng2023flexgen, kwon2023efficient, dao2022flashattention}, speculative decoding~\cite{leviathan2023fast}, and QoE-aware inference~\cite{liu2024andes}, all of them solely optimize for the generation inside the cloud datacenter before the tokens are generated but does not include the delivery of the tokens. As long as the user is following the newest generated content of the chatbot and packets get lost in the middle, a long stall might still happen and damage user experience.

% \hanchen{Need to talk about LLM works.}

% Besides FEC, there have also been error concealment~\cite{h264_error, kang2022error} and neural-network-enhanced loss resilience methods~\cite{cheng2023grace, li2023reparo} that utilize inter-frame and intra-frame to recover the lost part of the frame. These methods put extra computational requirements on the client end. In the case of token streaming, predicting the token reliably requires running a language model on the client device and is orthogonal to our solution.

\subsection{Unique Properties and Challenge}
\label{subsection:unique properties}
This {\em token streaming} problem bears some resemblance to the traditional streaming problems, but they have several distinct characteristics.
% Compared with the previous audio and video streaming problems, we characterize the following properties for our token streaming problem under unstable networks:
\begin{itemize}
    \item Unlike video/audio streaming, increasing the bitrate for transmission does not increase rendering quality.
    \item Sequence needs to be strictly maintained so that one lost token could block future rendering. In contrast, video/audio may skip frame~\cite{salsify} under loss. 
    \item The minimal bandwidth requirement for the network is much lower since text data are much smaller than audio (150Kbps) and video (>1Mbps).
\end{itemize}

Given these properties, we believe that in order to render tokens smoothly even under unstable networks, a robust token streaming method should prevent the problem of lost packet blocking subsequent arrived tokens rendering. For this to happen, a moderate increase in total sent data is acceptable. The main challenge we face is intermittent connectivity, while the average bandwidth could stay over 100Kbps even when the network is unstable~\cite{reboost}. Moreover, we could add redundancy without suffering from quality drops. 
\section{\name Design}

We propose \name, a novel transmission scheme that puts ``unacked tokens'' in the outgoing packets that contain the newly generated tokens, in order to reduce client token streaming stalls under an unstable network.

The key idea of \name is to prepare packets so that the client can render each received packet independently. That is, each packet contains not only newly generated tokens but also enough information to render these new tokens. Thus, for each arrived new token, the client is able to render the current sentence all the way until this new token and would not potentially be blocked by any previously lost token. With this mechanism, retransmission-incurred stalls can be greatly reduced, especially under unstable network where round-trip-time (RTT) can be long.

\subsection{Illustrative Example}
\label{protocol:example}
We showcase one example case of \name comparing with TCP and naive redundancy in Fig. \ref{fig:comparison}. In this scenario, we are streaming a response of ``You can go to watch...'' and the client has received tokens to render ``You can''. The LLM model now outputs ``go'', ``to'', and ``watch'' sequentially with time gap in between. We assume the unstable network will drop the first two packets sent and successfully deliver the following packets. 

In the current TCP retransmission logic, after the two packets containing tokens \#1 and \#2 are lost, token \#3 successfully arrives at the receiver. The receiver sends an ACK back to the sender, and the sender retransmits the first two tokens to the receiver again to finally render the whole sentence.

In the second case, we added one duplication packet for each outgoing token. However, if the first two packets are lost, the first token is still lost during transmission. Although tokens \#2 and \#3 successfully arrive at the receiver, their rendering is blocked by token \#1. The client still has to wait for the retransmission of token \#1 to render the sentence.

In the last column, we demonstrate how our method could reduce the stall. After the first token is sent, the following packets will incorporate token \#1 into the payload. Similarly token \#2 is contained in later packets. Thus, when the packet containing the new token \#3 arrives at the receiver, the receiver is able to render the whole sentence since the previous tokens needed to render \#3 are all contained in the packet. This greatly reduces the stall perceived on the client side. Moreover, the total size of sent data is similar to the original TCP and less than duplication since retransmission adds extra workload to the system.

\begin{algorithm}
\caption{\name Sender Logic}\label{alg:sender}
\begin{algorithmic}

\STATE $unacked \gets \text{empty}$

\WHILE{EOS Token not Acked}
\STATE {Sleep until newly generated $token$ or ($unacked\_buffer \neq empty$ and slept for 200ms) }
\STATE $unacked \gets unacked \setminus new\_acked\_tokens$

\IF{size of $unacked \leq$ max size of packet}
\STATE $packet \gets token$ concatenated with $unacked$ 
\ELSE
\STATE $remainingSize \gets$ max size of packet $-$ size of $token$
\STATE $partial\_unacked \gets \text{empty}$

\FOR{$i \gets 0 \textbf{ to } \text{len(unacked)}$}
        \IF{$\text{sizeof}(partial\_unacked) + $$\text{sizeof}(\text{unacked}[i])$$ < remaining\_size$}
            \STATE $ partial\_unacked.\text{append}(\text{unacked}[i])$
        \ELSE
            \STATE \textbf{Break}
        \ENDIF
    \ENDFOR
\STATE $packet \gets packet$ concatenated with $partial\_unacked$
\ENDIF

\STATE Send $packet$
\STATE Append $token$ into $unacked$ 
\ENDWHILE
\end{algorithmic}
\end{algorithm}
\vspace{-5pt}
\subsection{Specific Workflow}
We provide the pseudo-code for \name scheme in algorithm \ref{alg:sender} and algorithm \ref{alg:receiver}. The core logic of the sender is to maintain a unacked token buffer that contains the all tokens that were sent but have not received ACK. Every time a new token is generated, we visit this buffer to put as many unacked tokens as maximal size allows into the packet with the new token to prepare for the rendering of new token. The receiver follows a simpler logic. After receiving a packet, it extracts the tokens inside, renders the tokens if all previous tokens have arrived, and sends back ACKs to the sender. This scheme could be implemented inside RTP~\cite{rtp} or QUIC~\cite{rfc9000}, since we allow sent data to never arrive at client side.

\begin{algorithm}
\caption{\name Receiver Logic}\label{alg:receiver}

\begin{algorithmic}

\STATE $acked\_list \gets$ empty list

\WHILE{EOS was not received}
\STATE {Sleep until receives $Packet$}
\STATE $token\_list \gets$ tokens in packet
\FOR{$token$ in $token\_list$}
\IF{$token$ not acked before}
\STATE $acked\_list \gets$ $acked\_list$ append $token$ 
\ENDIF
\ENDFOR
\STATE Render new tokens according to $acked\_list$

\STATE Send ACK for $packet$
\ENDWHILE
\end{algorithmic}
\end{algorithm}

\subsection{Limitation}
\label{sec:limitation}
Admittedly, our method has limitations. The most notable limitation is that sometimes not all unacked tokens could fit into one packet. Mathematically, this happens when:
\vspace{-8pt}

\[
G \times (T - 1) \leq 2 \times RTT + L,
\] 
where $G$ is the time gap between two generated tokens, $T$ is the maximum number of tokens in one packet, RTT is the round trip time, and $L$ is the packet loss time.

As measured in~\cite{measurement_gpt}, most of the time current GPT4 throughput is around 10 tokens/s. Since human normal reading speed is 4-6 words per second (fast reading speed is 8-15 words per second)~\cite{speed_reading}, LLM service providers have less incentive to purchase more GPUs to drastically increase throughput beyond this level when more users contend for the computing resources in the future. On the other hand, 
% although the current packet size for one token is from 70 to 300 bytes depending on the application, 
it is possible to reuse the meta-data inside each packet to save spaces for the token data. In worst case, if the unacked tokens cannot all fit in the new packet, we only add the earlier unacked tokens into the packet in order to make sure that some tokens can still be independently rendered after arrival.

\section{Evaluation}
\subsection{Setup}
We measured our performance in an end-to-end Python simulation platform, which takes in generation timestamps of tokens, network channel parameters, as well as a sender/receiver pair, and outputs rendering time for each token. We report two metrics in our report: redundancy rate and stall ratio. Redundancy rate is the extra amount of data sent (including retransmission) besides the original tokens. We borrowed the idea of stall ratio from video streaming~\cite{cheng2023grace, stall} and defined it to be the amount of time that client side is waiting for tokens to render when stall is longer than 200ms. As discussed in \S\ref{subsection:unique properties}, an ideal method should prioritize reducing stall ratio and try to have lower redundancy rate afterwards.

We implemented a pair of \name sender/receiver in $\sim70$ lines of code and set the maximal number of tokens in one packet to be 10. We also implemented a group of packet duplication senders, which send multiple copies of every packet to prevent loss (similar to FEC in video streaming). 

In our experiments, We set input generation sequence to generate new token every 100ms as the current GPT-4 generation throughput is around 10 tokens/s~\cite{measurement_gpt}. For each setting, we run 30 30-second sessions.

\begin{figure}[t]
\centering
     \includegraphics[width=.99\linewidth]{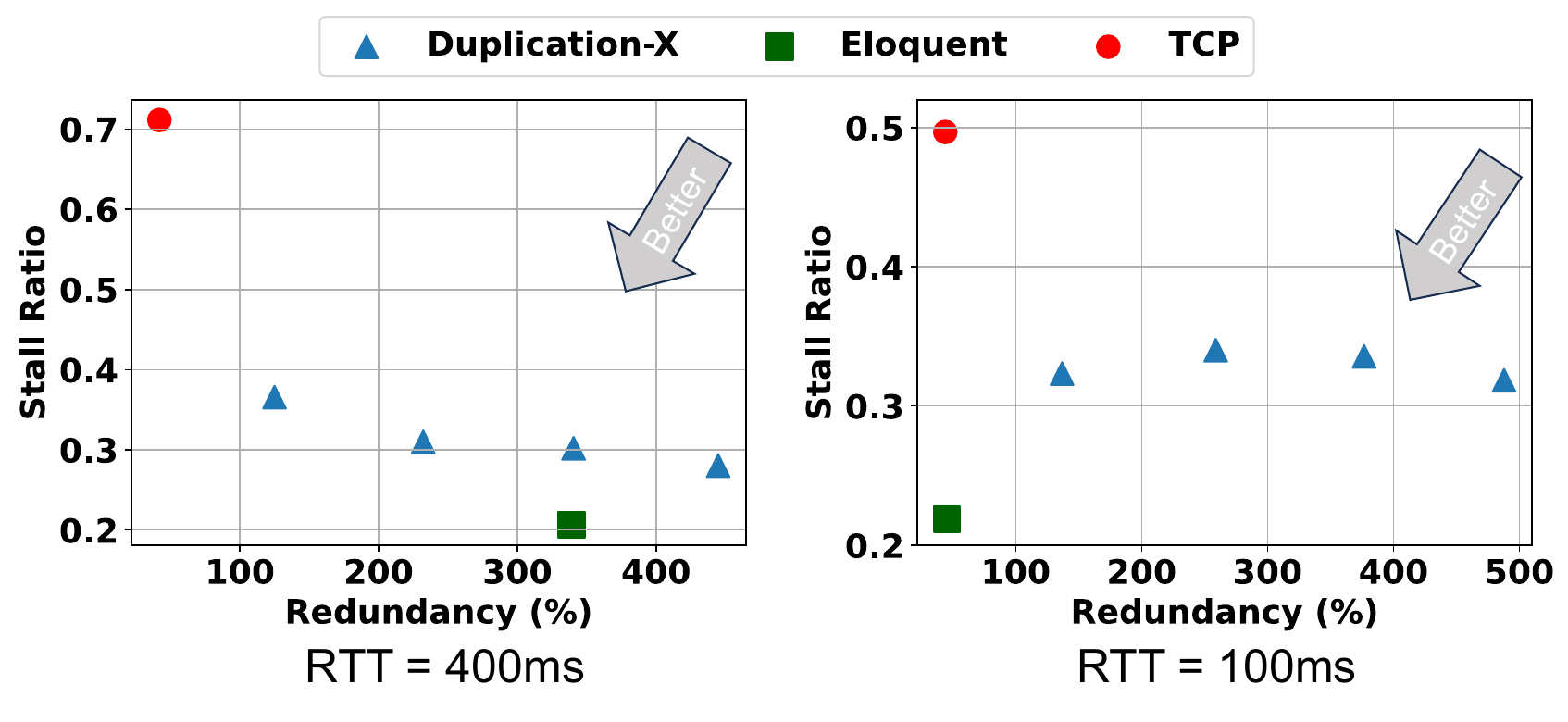}
    \tightcaption{\name reduces stall ratio under 15\% loss rate. Duplication-X are with rate 2x, 3x, 4x, 5x from left to right.} 
    % \vspace{-5pt}
    \label{fig:eval1}
\end{figure}
\begin{figure}[t]
\centering
     \includegraphics[width=.99\linewidth]{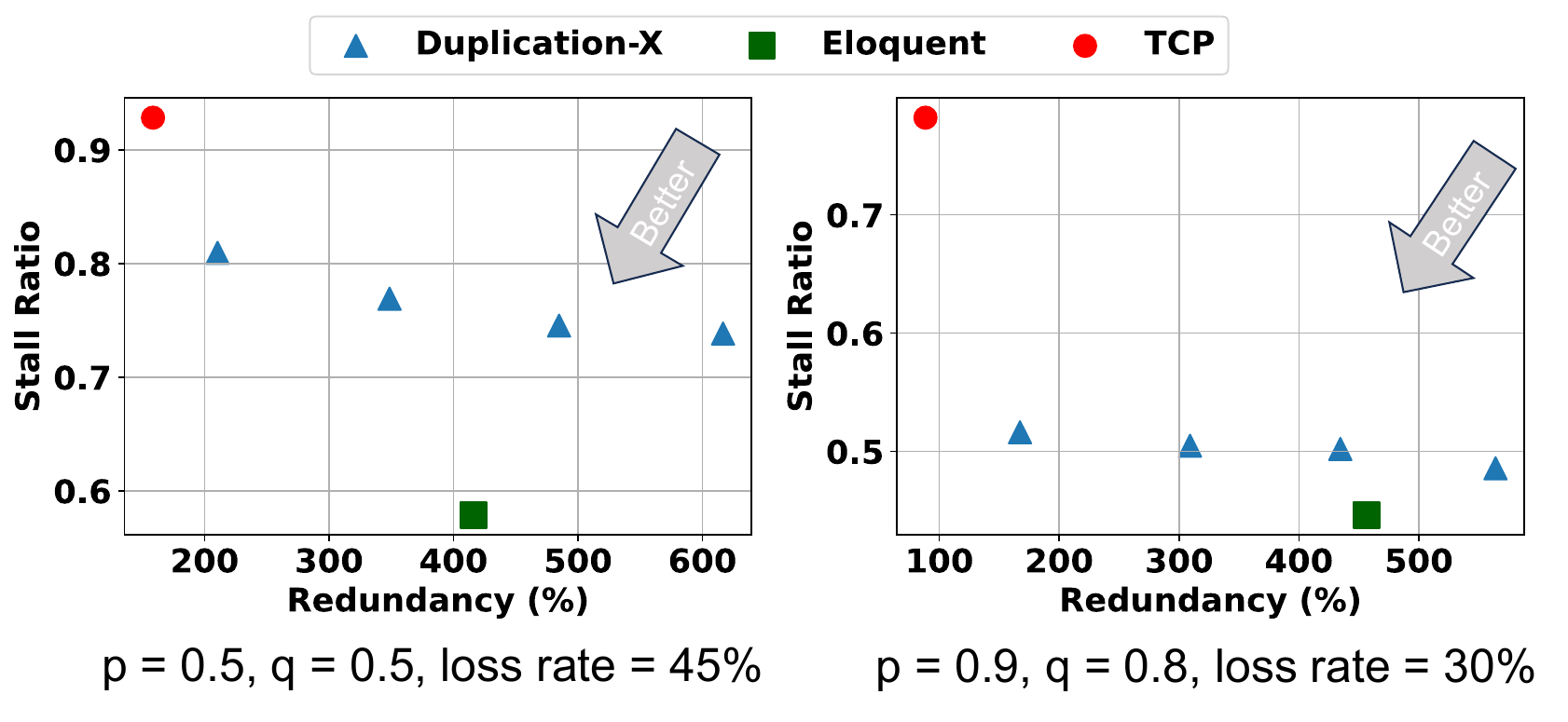}
\vspace{-5pt}
    \tightcaption{\name performs well under various loss patterns
} 
\vspace{-5pt}
    \label{fig:eval2}
\end{figure}

\subsection{Experiment Results}
In this part, we utilize a two-state Markov model~\cite{loss_model, loss_model2} to simulate unstable network conditions like wireless connections. We set two state: lossy state, when packets are lost with 90\% possibility; and good states, when no packets are lost. Each state lasts for 100ms. There are two parameters p and q: p is the probability to transform from good state to lossy state and q is the probability to change in the opposite direction. On average, the channel enters lossy state for $\frac{100}{q}$ms after staying in good state for $\frac{100}{p}$ms. Our default case, where $p= 0.9$ and $q= 0.5$, has 200ms bad connections every 1000ms good connection on average. This gives a packet loss rate of 15\%, similar to what we observed in the measurement in \S\ref{emulated_measurement}.

We start with the default setting and set round trip time (RTT) to 400ms and 100ms in Fig. \ref{fig:eval1}. As shown on the left, \name is able to reduce stall ratio by 71.0\% compared with the original TCP baseline. This is because \name reduces the need for retransmission under packet loss. Comparing with packet duplication methods, \name is able to reduce stall by 31.6\% percent with similar redundancy rate. The right graph with 100ms RTT shows a similar result, when \name reduces stall ratio by 60.0\% compared with TCP and 31.3\% compared with duplication with less size.
% Notice that redundancy rate of \name is higher when RTT is longer. This is because it takes longer for sent packets to be acknowledged even without loss, which causes \name to put more unacked tokens in each packet. However, this extra caution is necessary, since delay cost of retransmission also increases with longer RTT.

To learn about how \name performs under different loss patterns, we fixed RTT to be 400ms and modified the loss parameters in Fig. \ref{fig:eval2}. The left graph shows results with $p = 0.5$ and $q = 0.5$. This network is more likely to enter lossy state than default. Since loss is more frequent, the stall ratios of all methods increased. But \name's improvements is more significant since packet loss are more frequent. On the right side we set $p$ to 0.9 and $q$ to 0.8. In this case, every lossy period lasts longer. As shown in the right hand side, although we still outperforms the baselines, our improvement decreased due the limitation mentioned in \S\ref{sec:limitation}. Since both loss time (L) and RTT are big, many packets could not contain all unacked tokens, and this damages \name's performance.
This potentially could be improved by reusing packet meta-data (font type, size...) to put more tokens in each packet. 

\subsection{Real Trace Experiement}
We utilized packet arrival traces collected from the emulation platform in \S\ref{emulated_measurement} to conduct end-to-end simulation under real-world loss patterns. We calculated whether the ith token packet was lost by TCP sequence number and used this data to determine packet loss in the simulation sender channel. 

\begin{figure}
\centering
     \includegraphics[width=\linewidth]{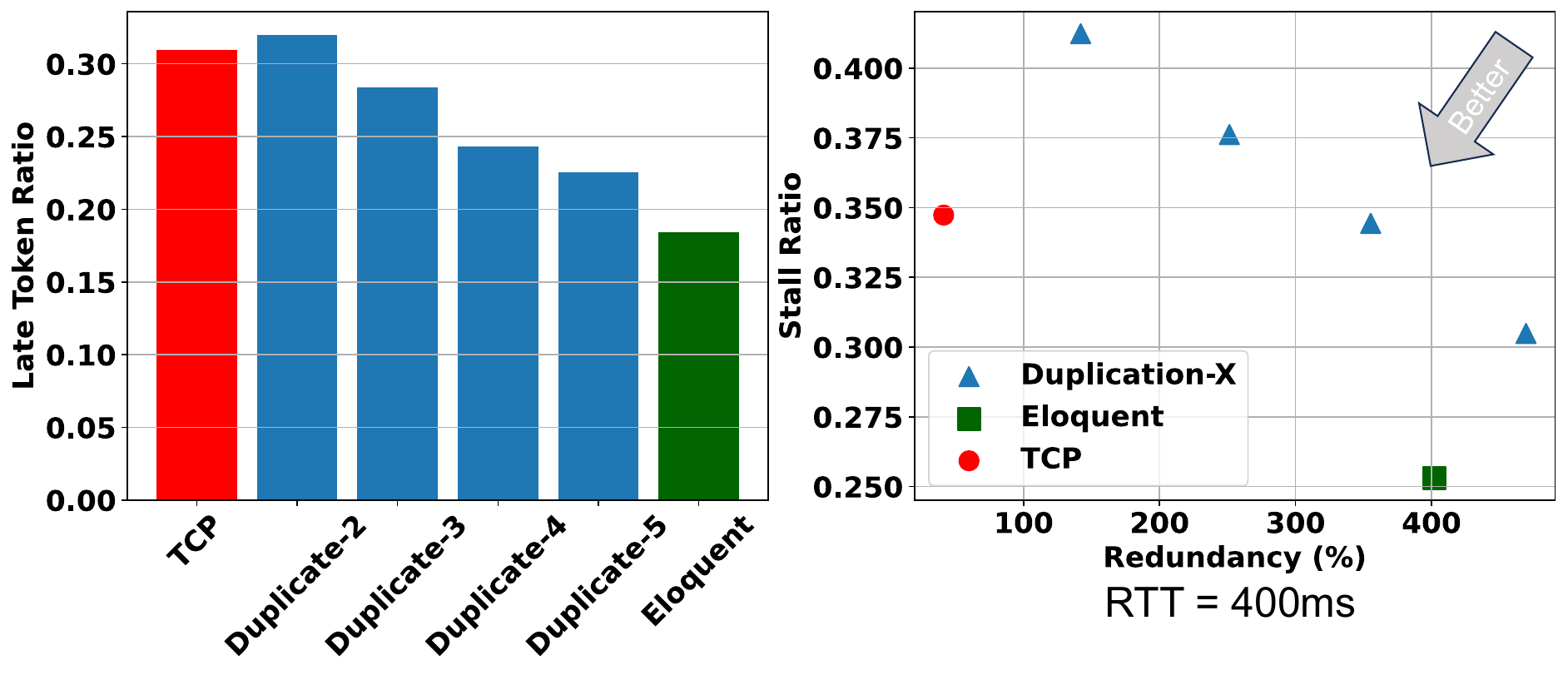}
    \vspace{-10pt}
    \tightcaption{\name reduces stall ratio and late tokens greatly in Real Trace Driven Experiement}
               \vspace{-3pt} 

    \label{fig:eval3}
\end{figure}

As shown in Fig \ref{fig:eval3}, under 400ms RTT, \name is able to reduce stall ratio by 27.4\% compared with TCP and 18.0\% compared with packet duplication with less size. Moreover, it reduces the proportion of late tokens (delay since generation over 400ms) by 40.4\% over TCP, providing more real-time responses with no modification to computation. 

\section{Future Directions}
\label{sec:discussion}

\subsection{Practical implementation} 
This paper only intends to demonstrate the promising improvement of a new transmission scheme over unstable network. More work is still needed before implementing \name in practice. 

To start with, since \name does not require all sent bytes to receive at the receiver, \name should be implemented within one of the transport protocols based on UDP. 
For reference, \name could be implemented in QUIC for the stream of generated text as its loss recovery method, or on top of RTP similar to audio/video streaming. Different from the default TCP method that maintains the unacked buffer inside the operating system, \name maintains the unacked buffer inside application space. 

\subsection{Quality of Experience for Token Streaming}
There could be further effort on establishing the quality of experience (QoE) model for token streaming. Prior results in video QoE could provide insight on new system optimizaiton. For example, in video streaming it is established that lower quality frames are better than having stalls~\cite{cheng2023grace,qoe_general}. But in LLM Chatbot services people might be willing to suffer from rendering stall in order to receive more sophisticated answers. Moreover, users might have preference for a single long stall over frequent stalls. If this holds, then maintaining a token buffer at client side could help bring better experience. Building a better QoE model for LLM inference potentially improve our tradeoff curve for inference systems. 

\subsection{QoE-aware LLM Inference}
Prior works including Andes~\cite{liu2024andes} and Google's Bard~\cite{bard} maintains a buffer at client side to gradually release text at human reading speed. This damages the real-timeness of response but increases smoothness while alleviating the pressure on system scheduling. \name could be combined with these systems to selectively enter redundancy mode when the buffer is close to empty since a potential packet loss would cause more damage during that time. 

Moreover, we could also combine different inference systems. For example, VLLM optimizes throughput by increasing the batch size but may incur generation stall~\cite{kwon2023efficient}. While Sarathi-Serve has smoother generation with potentially lower throughput~\cite{agrawal2024taming}. With a user-side token buffer, we could optimize throughput while the buffer has many tokens and switch to methods like Sarathi-Serve to prevent generation stall when buffer is shallow.

\subsection{Exposing user state in LLM serving}
\name targets the problem of token streaming after the tokens are generated by the data center. There could also be works that sends signals to data center to improve token generation. For example, when users switches tab, a signal could be transmitted back to the data center to deprioritize smoothness since users are not actively reading the content. On the otherhand, if the network is detected to be lossy, the client could try to switch to buffering mode and notify the user that the content will be available later.
\section{Conclusion}
In conclusion, we identify the transmission problem in the token streaming pipeline for LLM Chatbots and reveal the degraded performance of current application under unstable network with a measurement study. To improve the novel token streaming problem, we propose \name, a novel transmission scheme that proactively puts unacked tokens into packets of newly generated tokens. Through simulation, \name reduces stall ratio by 71.0\% over TCP and 31.6\% over baselines. In the end, we point out future research directions optimizing LLM inference with components outside data centers in consideration to spark more discussion.

\section*{Acknowledgement}
We thank all the anonymous reviewers, Xu Zhang, Qizheng Zhang, and Zhengxu Xia for their constructive feedback. This project is supported by NSF CNS-2146496, CNS-1901466, CNS-2313190.

\clearpage
% %-------------------------------------------------------------------------------
\bibliographystyle{plain}
\bibliography{citations} 
% \bibliography{\jobname}
% \clearpage

% % \input{3-2-new.tex}

% \section{Appendix}
% \input{9-appendix}

\pagebreak

\end{document}